\documentclass[letterpaper, 10pt]{article}

\begin{document}

\twocolumn[
\begin{center}
{\Large\bf BCS Theory for Binary Systems with 2D Electrons}\\
\vspace{0.15in}

X. H. Zheng\\
\vspace{0.05in}
\small
\em Department of Pure and Applied Physics, The Queen's University of Belfast,
Belfast BT7 1NN, Northern Ireland \em
\end{center}

\small
\begin{quote}
MgB$_2$ is considered as a binary system with 2D electrons.  
The classic BCS theory is applied to this system.  The 
transition temperature ($T_c$) is found to be relatively high,
because 2D electrons are more capable of moving with the atoms,
on top of other special features of this system to enhance the 
electron-phonon interaction.  This system may also shed light on
the nature of superconductivity in cuprates.\\

\end{quote}]

\noindent
There is little doubt that MgB$_2$ ($T_c = 39$K) is a conventional 
superconductor \cite{Nagamatsu}.  The isotope effect is found to be 
strong, and the energy gap is also of the conventional type \cite
{Bud'ko, Karapetrov}.  Unsolved is the mystery about $T_c$ which seems 
to be either above or at the limit suggested by the theory of Bardeen, 
Cooper and Schrieffer (BCS) for phonon-mediated superconductivity 
\cite{Bud'ko}.  Therefore it is of current interest to apply the BCS 
theory to MgB$_2$, taking the special features of this compound into 
consideration, to see if a proper $T_c$ would arise. 

MgB$_2$ features alternating B and Mg rings \cite{Nagamatsu}.
Energy band calculation reveals that superconductivity in MgB$_2$ is 
driven by the 2D bounding $\sigma$ bands \cite{An}.  We consider a simple 
model of alternating layers of light and heavy atoms, to which electrons 
pertain (Fig,\ 1).  These electrons are more capable of moving with the 
atoms, compared with free electrons, to enhance the electron-phonon 
interactions.  These interactions may take place either within the 
same atom layer (intra-layer coupling), or between neighboring layers 
(inter-layer coupling).  The intra and inter-layer couplings may join 
force to raise $T_c$ still further.

Our simple model might help us to understand cup-rates, regardless of 
the nature of the pairing mechanism.  In cuprates atoms are also layered, 
with electrons pertaining to the atom layers \cite{Pickett}.  The intra 
and inter-layer couplings are in competition: if none dominate then both 
are suppressed.  This leads to a forbidden zone of doping, where the 
inter-layer coupling is stronger than the coupling within the heavy atom 
layers, but weaker than that within the light atom layers, so that 
superconductivity vanishes, reminiscent of the observation by Moodenbaugh
et al.\ in La$_{2-x}$Ba$_x$CuO$_4$ \cite{Moodenbaugh}.  Careful doping may 
leave singlet pairs inside the forbidden zone, triplet pairs outside: 
$p$-wave pairing is possible in the BCS theory, and indeed in a large 
number of candidate theories, where $p$-wave symmetry is not intrinsically 
favored \cite{Waldram}.  

We apply the BCS theory to our binary system.  The derivation is simple 
in principle but involved technically.  Second quantization is actually 
based on the assumption that, despite its statistical nature, the 
displacement of a particle in a crystal is the exact replica of the 
displacement of its neighbors, save a phase factor, which serves as
a parameter to categorize particle displacement into groups known as 
Fourier components.  The operation to add or take away a component is
symbolized by the creation or destruction operator.  In this respect 
there is little difference between normal metals and our binary system.  
Phonons (atom displacements) have long been treated in complex systems 
\cite{Ziman}.  We treat electrons with the understanding that proper 
wave functions can be found to reflect the physics of 2D electrons: 
they pertain to the atom layers, and their wave functions overlap with 
those in neighboring layers.  When $T > 0$ we find the following 
ensemble average of electron energy 
\begin{eqnarray}
& & W = \sum^2_{i = 1}2\sum _{\bf k}\epsilon _{\bf k}
			 \left[f^{(i)}_{\bf k} + h^{(i)}_{\bf k}
			 (1 - 2f^{(i)}_{\bf k})\right]\nonumber\\
& & - \sum^2_{i,j = 1}\sum _{\bf k, k'}V^{(i,j)}_{\bf k,k'}
       \left[h^{(i)}_{\bf k}(1 - h^{(i)}_{\bf k})\right]^{1/2}
	   (1 - 2f^{(i)}_{\bf k})\nonumber\\
& & \times\left[h^{(j)}_{\bf k'}(1 - h^{(j)}_{\bf k'})\right]^{1/2}
       (1 - 2f^{(j)}_{\bf k'})
\label{eq:W}
\end{eqnarray}
\begin{picture}(228,160)(5,0)
\put(18,0){\framebox(228,160)}
\multiput(36,  0)(19,0){11}{\line(0,1){2}}
\multiput(36,158)(19,0){11}{\line(0,1){2}}
\multiput( 18, 16)(0,16){ 9}{\line(1,0){2}}
\multiput(244, 16)(0,16){ 9}{\line(1,0){2}}

\put( 12,-15){\sffamily .130}
\put( 64,-15){\sffamily .132}
\put(122,-15){\sffamily .134}
\put(179,-15){\sffamily .136}
\put(232,-15){\sffamily .138}

\put(5,  0){\sffamily  0}
\put(5, 77){\sffamily 10}
\put(5,153){\sffamily $V$}

\put(105,-30){VALENCY(Z)}

\thicklines
\qbezier(18,82.67)(31.68,77.80)(49.92,74.02)
\put(49.92,0){\line(0,1){74.02}}
\put(186.7,0){\line(0,1){42.43}}
\qbezier(186.7,42.43)(214.06,40.696)(246,38.67)
\multiput(18,23.696)(7.6,-0.2461333){31}{\circle*{1.5}}

\put(187,50){2D electrons}
\put(92,32){free electrons}

\thinlines
\multiput(95.82,142.82)(24,0){2}{\circle*{3}}
\multiput(95.82,102.82)(24,0){2}{\circle*{3}}

\multiput(83,130)(24,0){2}{\circle*{3}}
\multiput(83, 90)(24,0){2}{\circle*{3}}

\multiput(83,130)(12.82,12.82){2}{\line(1,0){24}}
\multiput(83, 90)(12.82,12.82){2}{\line(1,0){24}}

\multiput(83,130)(24, 0){2}{\line(1,1){12.82}}
\multiput(83, 90)(24, 0){2}{\line(1,1){12.82}}

\multiput(95.82,122.82)(24,0){2}{\circle{4}}
\multiput(95.82, 82.82)(24,0){2}{\circle{4}}

\multiput(83,110)(24,0){2}{\circle{4}}
\multiput(83, 70)(24,0){2}{\circle{4}}

\multiput(85,110)(12.82,12.82){2}{\line(1,0){20}}
\multiput(85, 70)(12.82,12.82){2}{\line(1,0){20}}

\multiput(84.41,111.41)(24,0){2}{\line(1,1){10}}
\multiput(95.82,124.82)(24,0){2}{\line(0,1){18}}

\multiput(84.41, 71.41)(24,0){2}{\line(1,1){10}}
\multiput(83,112)(24,0){2}{\line(0,1){18}}

\multiput(95.82, 84.82)(24,0){2}{\line(0,1){36}}
\multiput(83, 72)(24,0){2}{\line(0,1){36}}

\multiput(66,70)(0,60){2}{\line(1,0){ 6}}
\multiput(69,70)(0,38){2}{\line(0,1){22}}

\put(082,138){$a$}
\put(105,145){$b$}
\put(067,098){$c$}

\put(125,120){1}
\put(125,100){2}

\put(143, 82.82){\line( 0,1){60}}

\put(143,140.82){\line( 1,0){ 1.8772}}
\put(143,138.82){\line( 1,0){ 3.7082}}
\put(143,136.82){\line( 1,0){ 5.4479}}
\put(143,134.82){\line( 1,0){ 7.0534}}
\put(143,132.82){\line( 1,0){ 8.4853}}
\put(143,130.82){\line( 1,0){ 9.7082}}
\put(143,128.82){\line( 1,0){10.6921}}
\put(143,126.82){\line( 1,0){11.4127}}
\put(143,124.82){\line( 1,0){11.8523}}
\put(143,122.82){\line( 1,0){12.0000}}
\put(143,120.82){\line( 1,0){11.8523}}
\put(143,118.82){\line( 1,0){11.4127}}
\put(143,116.82){\line( 1,0){10.6921}}
\put(143,114.82){\line( 1,0){ 9.7082}}
\put(143,112.82){\line( 1,0){ 8.4853}}
\put(143,110.82){\line( 1,0){ 7.0534}}
\put(143,108.82){\line( 1,0){ 5.4479}}
\put(143,106.82){\line( 1,0){ 3.7082}}
\put(143,104.82){\line( 1,0){ 1.8772}}
\put(143,100.82){\line(-1,0){ 1.8772}}
\put(143, 98.82){\line(-1,0){ 3.7082}}
\put(143, 96.82){\line(-1,0){ 5.4479}}
\put(143, 94.82){\line(-1,0){ 7.0534}}
\put(143, 92.82){\line(-1,0){ 8.4853}}
\put(143, 90.82){\line(-1,0){ 9.7082}}
\put(143, 88.82){\line(-1,0){10.6921}}
\put(143, 86.82){\line(-1,0){11.4127}}
\put(143, 84.82){\line(-1,0){11.8523}}
\put(143, 82.82){\line(-1,0){12.0000}}

\put(168, 82.82){\line( 0,1){60}}

\put(168,142.82){\line( 1,0){12.0000}}
\put(168,140.82){\line( 1,0){11.8523}}
\put(168,138.82){\line( 1,0){11.4127}}
\put(168,136.82){\line( 1,0){10.6921}}
\put(168,134.82){\line( 1,0){ 9.7082}}
\put(168,132.82){\line( 1,0){ 8.4853}}
\put(168,130.82){\line( 1,0){ 7.0534}}
\put(168,128.82){\line( 1,0){ 5.4479}}
\put(168,126.82){\line( 1,0){ 3.7082}}
\put(168,124.82){\line( 1,0){ 1.8772}}
\put(168,120.82){\line(-1,0){ 1.8772}}
\put(168,118.82){\line(-1,0){ 3.7082}}
\put(168,116.82){\line(-1,0){ 5.4479}}
\put(168,114.82){\line(-1,0){ 7.0534}}
\put(168,112.82){\line(-1,0){ 8.4853}}
\put(168,110.82){\line(-1,0){ 9.7082}}
\put(168,108.82){\line(-1,0){10.6921}}
\put(168,106.82){\line(-1,0){11.4127}}
\put(168,104.82){\line(-1,0){11.8523}}
\put(168,102.82){\line(-1,0){12.0000}}
\put(168,100.82){\line(-1,0){11.8523}}
\put(168, 98.82){\line(-1,0){11.4127}}
\put(168, 96.82){\line(-1,0){10.6921}}
\put(168, 94.82){\line(-1,0){ 9.7082}}
\put(168, 92.82){\line(-1,0){ 8.4853}}
\put(168, 90.82){\line(-1,0){ 7.0534}}
\put(168, 88.82){\line(-1,0){ 5.4479}}
\put(168, 86.82){\line(-1,0){ 3.7082}}
\put(168, 84.82){\line(-1,0){ 1.8772}}

\put(134,65){$\psi^{(1)}_{\bf k}$}
\put(163,65){$\psi^{(2)}_{\bf k}$}

\end{picture}

\vspace{1.2cm}
\noindent 
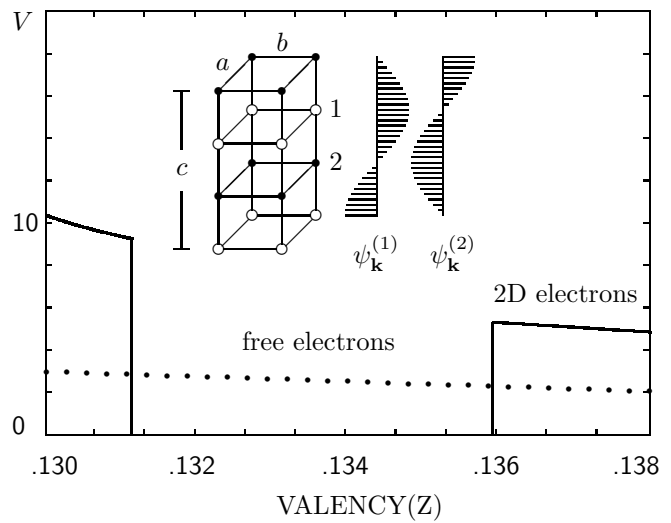
Fig.\ 1 Crystal of light (wt.\ $M_1$ in layer 1, open circles)
and heavy (wt.\ $M_2$ in layer 2) atoms, $a$, $b$ and $c$ are lattice
constants, $M_1/M_2 = 0.7$, $a = b = 0.5c$.  The anti-nodes of 
$\psi^{(1)}_{\bf k}$ and $\psi^{(2)}_{\bf k}$ are in layer 1 and 2,
respectively.  Values of $V$ are in an arbitrary unit.\\
Here {\bf k} is the wave vector to specify the electron state, 
$\epsilon _{\bf k}$ state energy relative to the Fermi level, 
$h_{\bf k}$ occupation probability of state {\bf k}, $f^{(i)}_{\bf k}$ 
excitation probability of the ground pair at states {\bf k} 
and $-{\bf k}$, and $i$ and $j$ indicate atom layers.  Following
BCS \cite{BCS} we associate ${\bf k}$ and $-{\bf k}$ with spin 
$\uparrow$ and $\downarrow$ respectively unless stated otherwise.  
We have
\begin{equation}
V^{(i,j)}_{\bf k,k'} = \sum^6_{l = 1}
\frac{2 \hbar\omega _{\mbox{\boldmath$\kappa$}l}
\mathcal{M}^{(i,j)2}_l({\bf k, k'})}
{(\hbar\omega _{\mbox{\boldmath$\kappa$}l})^2 
- (\epsilon _{\bf k} - \epsilon _{\bf k'})^2}
\label{eq:V_kk'}
\end{equation}
where $\omega _l$ is the phonon frequency, {\mbox{\boldmath$\kappa$} = 
${\bf k - k'}$} phonon vector, $l$ identifies phonon branches.  We have 2
atoms in the primitive cell, giving 6 branches in each polarization.  
However, transverse phonons do not interact with electrons \cite{Ziman} 
so that Eq.\ (\ref{eq:V_kk'}) involves just 6 branches (3 of them optic).  
The matrix element $\mathcal{M}^{(i,j)}_l$ measures the strength of 
electron-phonon interaction, which couples electrons as long as their 
wave functions overlap: the coupling may take place either within the 
same atom layer ($i = j$) or between neighboring layers ($i \neq j$).  
We have $\mathcal{M}^{(1,2)}_l = \mathcal{M}^{(2,1)}_l$, so that 
$V^{(1,2)}_{\bf k,k'} = V^{(2,1)}_{\bf k,k'}$.   

In the binary system the entropy of the electron ensemble is
\begin{displaymath}
S = -4k_B\sum _{\bf k}\Bigg[
\left(\frac{1}{2}f^{(1)}_{\bf k} + \frac{1}{2}f^{(2)}_{\bf k}\right)
\ln\left(\frac{1}{2}f^{(1)}_{\bf k} + \frac{1}{2}f^{(2)}_{\bf k}\right)
\end{displaymath}
\begin{equation}
+ \left(1 - \frac{1}{2}f^{(1)}_{\bf k} - \frac{1}{2}f^{(2)}_{\bf k}\right)
\ln\left(1 - \frac{1}{2}f^{(1)}_{\bf k} - \frac{1}{2}f^{(2)}_{\bf k}\right)
\Bigg]
\label{eq:S}
\end{equation}
where the factor 4 indicates the degree of degeneracy: at the same 
energy level there can be 2 electrons in each of the 2 atom layers.  
We have assumed the same density of states in both atom layers, 
otherwise the factor 1/2 in Eq.\ (\ref{eq:S}) must be adjusted.  
We minimize the free energy $F = W - TS$ with respect to 
$h^{(i)}_{\bf k}$ and find
\begin{equation}
\sum _{\bf k'}V^{(1,1)}_{\bf k,k'}
\frac{\Delta _1(1 -2f^{(1)}_{\bf k'})}{2\left(\Delta^2_1 
+ \epsilon^2_{\bf k'}\right)^{1/2}}
+\sum _{\bf k'}V^{(1,2)}_{\bf k,k'}
\frac{\Delta _2(1 -2f^{(2)}_{\bf k'})}{2\left(\Delta^2_2 
+ \epsilon^2_{\bf k'}\right)^{1/2}} = \Delta _1
\label{eq:self_T_1}
\end{equation}
\begin{equation}
\sum _{\bf k'}V^{(1,2)}_{\bf k,k'}
\frac{\Delta _1(1 -2f^{(1)}_{\bf k'})}{2\left(\Delta^2_1 
+ \epsilon^2_{\bf k'}\right)^{1/2}}
+\sum _{\bf k'}V^{(2,2)}_{\bf k,k'}
\frac{\Delta _2(1 -2f^{(2)}_{\bf k'})}{2\left(\Delta^2_2 
+ \epsilon^2_{\bf k'}\right)^{1/2}} = \Delta _2
\label{eq:self_T_2}
\end{equation}
which are counterparts of the BCS self-consistent equation, $\epsilon
_{\bf k}/(\Delta^2_i + \epsilon^2_{\bf k})^{1/2} = 1 - 2h^{(i)}_{\bf k}$
\cite{BCS}.  We also minimize $F$ with respect to $f^{(i)}_{\bf k}$ and find 
\begin{equation}
\left(\Delta^2_i + \epsilon^2_{\bf k}\right)^{1/2} = k_BT\ln\left[\frac{2 - 
f^{(1)}_{\bf k} - f^{(2)}_{\bf k}}{f^{(1)}_{\bf k} + f^{(2)}_{\bf k}}\right]
\label{eq:E_i}
\end{equation}
where $i = 1$ or 2, i.e.\ $\Delta _1 = \Delta _2 = \Delta$, which leads 
through Eqs.\ (\ref{eq:self_T_1}), (\ref{eq:self_T_2}) and the familiar BCS 
algorithm \cite{BCS} to
\begin{eqnarray}
  V_{11}\sum _{\bf k}\frac{1 - 2f^{(1)}_{\bf k}}
  {2\left(\Delta^2 + \epsilon^2_{\bf k}\right)^{1/2}}
+ V_{12}\sum _{\bf k}\frac{1 - 2f^{(2)}_{\bf k}}
  {2\left(\Delta^2 + \epsilon^2_{\bf k}\right)^{1/2}} 
= 1\label{eq:sol_1}\\
  V_{12}\sum _{\bf k}\frac{1 - 2f^{(1)}_{\bf k}}
  {2\left(\Delta^2 + \epsilon^2_{\bf k}\right)^{1/2}}
+ V_{22}\sum _{\bf k}\frac{1 - 2f^{(2)}_{\bf k}}
  {2\left(\Delta^2 + \epsilon^2_{\bf k}\right)^{1/2}}
= 1\label{eq:sol_2}
\end{eqnarray}
{\bf k} runs over a range where $-\hbar\omega < \epsilon _{\bf k} < 
\hbar\omega$, $\omega$ being the average phonon frequency, $V_{ij} = 
\langle V^{(i,j)}_{\bf k, k'}\rangle _{AV}$ matrix element averaged 
over the above range of ${\bf k}$.  We find from Eqs.\ (\ref{eq:sol_1}) 
and (\ref{eq:sol_2}) that
\begin{eqnarray}
\sum _{\bf k}\frac{1 - 2f^{(1)}_{\bf k}}{2\left(\Delta^2 
  + \epsilon^2_{\bf k}\right)^{1/2}}
  = \frac{V_{12} - V_{22}}{V^2_{12} - V_{11}V_{22}}\label{eq:sol_3}\\
\sum _{\bf k}\frac{1 - 2f^{(2)}_{\bf k}}{2\left(\Delta^2 
  + \epsilon^2_{\bf k}\right)^{1/2}}
  = \frac{V_{12} - V_{11}}{V^2_{12} - V_{11}V_{22}}\label{eq:sol_4}
\end{eqnarray}
We have assumed $M_2 > M_1$ (Fig.\ 1) so that $V_{22} < V_{11}$ (to be
justified later).  When $V_{22} < V_{12} < (V_{11}V_{22})^{1/2}$, the
right hand side of Eq.\ (\ref{eq:sol_3}) becomes negative, which means
$f^{(1)}_{\bf k} > 1/2$ must hold at least for some states.  This not
only violates the nature of $f^{(1)}_{\bf k}$ but also leads to a
contradiction: $\Delta$ may become so large that Eqs.\ (\ref{eq:self_T_1}) 
and (\ref{eq:self_T_2}) cannot balance.  Similarly, we find from Eq.\ 
(\ref{eq:sol_4}) that $(V_{11}V_{22})^{1/2} < V_{12} < V_{11}$ is not 
allowed.  Apparently 
\begin{equation}
V_{22} < V_{12} < V_{11}
\label{eq:forbidden}
\end{equation}
is a forbidden zone: if the inter-layer coupling is stronger than the
coupling within the heavy atom layers, but weaker than that within the 
light atom layers, then superconductivity is suppressed. Outside this 
forbidden zone we can add Eqs.\ (\ref{eq:sol_3}) and (\ref{eq:sol_4}) 
together and find through Eq.\ (\ref{eq:E_i}) that
\begin{equation}
\int^{\hbar\omega}_0\!\!\!\frac{d\epsilon}{\left(\Delta^2 
+ \epsilon^2\right)^{1/2}}\tanh\left[\frac{\left(\Delta^2 
+ \epsilon^2\right)^{1/2}}{2k_BT}\right] = \frac{1}{N(0)V}
\label{eq:sol_6}
\end{equation}
where the summation over {\bf k} has been converted into integration, $N(0)$
is the density of states at the Fermi surface.  Eq.\ (\ref{eq:sol_6}) is 
formally identical to the familiar BCS equation \cite{BCS} with
\begin{equation}
V = \frac{V^2_{12} - V_{11}V_{22}}
         {V_{12} - (V_{11} + V_{22})/2}
\label{eq:V}
\end{equation}
which is the counterpart of the BCS average matrix element \cite{BCS}.  

Now we evaluate $V^{(i,j)}_{\bf k,k'}$ in Eq.\ (\ref{eq:V_kk'}).  At 
this point we have to be more specific about the electron wave functions.  
For simplicity, we consider super-positions of two Bloch functions:
\begin{eqnarray}
\psi^{(1)}_{{\bf k}, \sigma}\propto\exp[i(k_xx + k_yy)]\cos(\pi z/c)
\label{eq:psi_1}\\
\psi^{(2)}_{{\bf k}, \sigma}\propto\exp[i(k_xx + k_yy)]\sin(\pi z/c)
\label{eq:psi_2}
\end{eqnarray}
where ${\bf k} = {\bf x}k_x + {\bf y}k_y$ is a 2D wavevector, $\sigma$ 
spin, $c$ lattice constant in the $c$-direction.  Both $\psi^{(1)}$ and 
$\psi^{(2)}$ have the same Bloch energy, and $\sigma = \,\uparrow$ or 
$\downarrow$, so that our electron wave functions are of quadruple 
degeneracy, consistent with the factor 4 in Eq.\ (\ref{eq:S}).  
Furthermore $\psi^{(1)}$ and $\psi^{(2)}$ are orthogonal to each other: 
they are proper base functions for second quantization. 

Both $\psi^{(1)}$ and $\psi^{(2)}$ are traveling waves in the $a$-$b$ 
plane but standing waves in the $c$-direction: they are mobile only in 
2D.  Apparently $\psi^{(1)}$ pertains to layer 1 (its anti-nodes are 
in that plane) whereas $\psi^{(2)}$ pertains to layer 2 (Fig.\ 1).  
It is implied that electron waves overlap with those in the neighboring 
layer, but not with those in the next neighboring layer (there is a node 
in between).  It appears that $\psi^{(1)}$ and $\psi^{(2)}$ are the 
leading and perhaps dominating terms of the electron configuration 
in a real binary system: dispersions arising from Eqs.\ (\ref{eq:psi_1}) 
and (\ref{eq:psi_2}) are similar to that in \cite{An}.

We average $V^{(i,j)}_{\bf k,k'}$ according to the text below Eq.\
(\ref{eq:sol_2}) and find
\begin{eqnarray}
V_{11} = \frac{C}{2}\int^{(8Z)^{-1/3}}_0\left[\frac{(A + B)^2}
{M_1} + \frac{(A - B)^2}{M_2}\right]\frac{\zeta^2d\zeta}{1 - \zeta^2}
\label{eq:V_11}\\
V_{12} = \frac{C}{2}\int^{(8Z)^{-1/3}}_0\left[\frac{B^2}{M_1} + 
\frac{B^2}{M_2}\right]\frac{0.65}{Z^{2/3}}\frac{a^3}{c^3}
\frac{d\zeta}{(1 - \zeta^2)^2}\,\,
\label{eq:V_12}
\end{eqnarray}
$V_{22}$ also is given by Eq.\ (\ref{eq:V_11}) with $M_1$ and $M_2$
interchanged.  Apparently $V_{11} > V_{22}$ when $M_1 < M_2$.  Here 
we have used the Debye phonon model for simplicity, $A$ and $B$ are 
values of the overlap integral function $F(x) = 3(\sin x -x\cos x)/x^3$, 
with $x = 3.84\alpha^{1/3}Z^{1/3}\zeta(1 - \zeta^2)^{1/2}$ and $3.84
\alpha^{1/3}[Z^{2/3}\zeta^2(1 - \zeta^2) + 0.65(a/c)^{4/3}]^{1/2}$, 
respectively, $C = 3Zm\alpha^2(\delta\mathcal{V})^2/N(0)\hbar\omega
\epsilon _F$, $Z$ is the average valency, $\epsilon _F$ the Fermi 
energy.  Eqs.\ (\ref{eq:V_11}) and (\ref{eq:V_12}) exist in the sense 
of the Cauchy principal value (used by Kuper to verify the BCS theory) 
\cite{Kuper}.  

The original BCS theory assumes a spherical Fermi surface, which may 
not be the case in binary systems \cite{An, Pickett}.  On the other 
hand, the states of 2D electrons form merely a cross-section of the 
Fermi sea: so long as this cross-section is circular, it makes little 
difference whether the Fermi sea is spherical or cylindrical, provided 
that $N(0)$ is of a proper local value.  Indeed, all the 2D calculations
over this circular cross-section parallel the calculations over the
spherical Fermi sea: no assumptions by BCS have been violated.

We let $M_1/M_2 = 0.7$ and $a = b = c/2$, and evaluate Eqs.\ 
(\ref{eq:V_11}) and (\ref{eq:V_12}) numerically (see Fig.\ 1).  
The forbidden zone is bordered by $Z = 0.1311$ and 0.1359: $V_{11}$, 
$V_{12}$ and $V_{22}$ are comparable when the valency is relatively 
small.  This arises from Eq.\ (\ref{eq:V_kk'}), where the matrix 
element $\mathcal{M}^{(1,2)}_l$ is proportional to the component 
wave number $\pi/c$ in Eqs.\ (\ref{eq:psi_1}) and (\ref{eq:psi_2}).  
In contrast $\mathcal{M}^{(1,1)}_l$ (or $\mathcal{M}^{(2,2)}_l$) are 
proportional to $k_x$ and $k_y$, which become smaller the smaller 
the Fermi surface ($Z$ smaller).  These lead to the following exotic 
properties:

\em Critical temperature\em: which can be relatively high for two 
reasons.  First, the intra and inter-layer couplings may join force 
to enlarge the energy gap.  For example, when $V_{12}\sim V_{11}$ 
in Eq.\ (\ref{eq:V}), $V\sim 2V_{11}$ holds in our binary system, 
compared with a normal metal or alloy where $V = V_{11}$.  This
is hardly a surprise, because in Eq.\ (\ref{eq:W}) the interaction
Hamiltonian always lowers the energy of the system, as long as
$V^{(i,j)}_{\bf k,k'} > 0$ and $f^{(i)}_{\bf k} < 1/2$, regardless
of whether the electron-phonon interaction takes place within the
same atom layer or between neighboring layers.  Second, electrons 
pertaining to atom layers are more capable of moving with the atoms.  
This can be seen from Eq.\ (\ref{eq:V_11}) where $V_{11}$ is 
proportional to $2A^2/M_1$ when $A = B$.  In a binary alloy with 
free electrons Eq.\ (\ref{eq:V_11}) also applies, with both $A + B$ 
and $A - B$ being replaced by $A$.  Therefore $V_{11}$ becomes 
proportional to $A^2/M_1$ when $M_1 = M_2$: free electrons are less 
capable of moving with the atoms, so that the electron-phonon 
interaction is weakened by half.  Indeed, in Fig.\ 1 the value of $V$ 
for 2D electrons is 3.35 times larger than that for free electrons
when $V_{12} = V_{11}$.  We use the Debye phonon model when 
deriving Eqs.\ (\ref{eq:V_11}) and (\ref{eq:V_12}), so that the 
contribution by optic phonons is somewhat overestimated.  However,
values of $V$ for free electrons are also overestimated, so that we
can still use the data in Fig.\ 1 to compare the energy gap in a 
normal metal or alloy with that in our binary system. 

\em Anomalous suppression of superconductivity\em: which is considered 
as one of the long-standing mysteries associated with high $T_c$ 
cuprates \cite{Tranquada}.  It was found that in La$_{2-x}$Ba$_x$CuO$_4$ 
$T_c \approx 25$K when $x = 0.09$ and $x = 0.15$, whereas $T_c\approx 5$K 
between these two maxima \cite{Moodenbaugh}.  According to energy band 
calculation, in LaBaCuO$_4$ the Fermi surface is crossed by the CuO band 
and two O$_z$ bands \cite{Pickett}: superconducting carriers are in atom 
layers which on average have two different masses.  Perhaps LaBaCuO$_4$ 
could be modeled as a binary system, which falls into the forbidden zone 
when the valancy is adjusted into a proper value through doping.

\em Spin triplet pairs\em: which may arise when spin singlet pairs fall 
into the forbidden zone.  Specifically, while $V_{12} < V_{11}$ holds for 
singlet pairs (inside the forbidden zone, suppressed), $V_{12} > V_{11}$ 
may hold for triplet pairs (allowed), because $V_{11}$ declines faster 
than $V_{12}$ when the pair symmetry changes, due to the stronger effect 
of the exchange term on $V_{11}$, which arises from intra-layer couplings, 
where electron waves overlap to a greater extent, compared with the 
inter-layer coupling.  It is interesting that magnetic excitations are 
observed in YBa$_2$Cu$_3$O$_8$ and Bi$_2$Sr$_2$CaCu$_2$O$_8$ \cite{Fong}.  
According to Yin et al.\ these excitations may not be attributed to the 
normal state properties of cuprates, because the excitation intensity 
decreases continuously with increasing temperature, and vanishes above 
$T_c$ \cite{Fong, Yin}.  Furthermore, magnetic excitations are observable 
only when doping is optimized \cite{Fong}.  Perhaps these cuprates may
also be modeled as a binary system, where careful doping may drive 
singlet pairs into the forbidden zone, leaving triplet pairs out.  It
is difficult to dope MgB$_2$, so that singlet pairs cannot be 
discriminated: the energy gap must be of the BCS type \cite{Karapetrov}.

In conclusion, the BCS theory may explain the relatively high transition
temperature of MgB$_2$ which, according to Bud'ko et al., nearly double
the previous record for a nonoxide and non-C60-based compound 
\cite{Bud'ko}.  We find that in a binary system with 2D electrons
$V$ can increase 2-3 times (Fig.\ 1).  According to Yin 
et al.\ many experimental observations in cuprates are commonly fit with a 
phenomenological model that derives from the original BCS theory, 
characterized by a gap equation corresponding to a presumed symmetry of the 
order parameter, with an adjustable dimensionless coupling constant, and a 
given Fermi surface \cite{Yin}.  We find that in our binary system the 
$p$-wave symmetry may arise in a natural manner from the BCS theory.  
This may help us to develop a microscopic theory for cuprates.  In this
letter we have used the original BCS theory, where damping and retardation
are ignored with negligible effect \cite{Schrieffer}.  This effect is 
likely to be similar, whether the superconductor is a normal metal, or 
a binary system with 2D electrons.  Therefore our major conclusion, i.e.\
$V$ can be 2-3 times larger in the later case, is likely to stand.  
In future we would like to improve our results by taking damping and 
retardation into consideration.\\
\\
{\bf Acknowledgments}:\\
The author thanks Professor D.\ G.\ Walmsley for helpful comments.

\end{document}